\documentclass[10pt,aps,prd,amsmath,amssymb,twoside,showpacs,nofootinbib,preprintnumbers]{revtex4}
\usepackage[T1]{fontenc}

\newcommand{\defeq}{:=}
\newcommand{\im}{\mathrm{i}}  

\newcommand{\C}{\mathbb{C}}
\newcommand{\xd}{\mathrm{d}}
\newcommand{\xD}{\mathcal{D}} 
 
\newcommand{\cH}{\mathcal{H}}

\newcommand{\be}{\begin{equation}}
\newcommand{\ee}{\end{equation}}
\newcommand{\bea}{\begin{eqnarray}}
\newcommand{\eea}{\end{eqnarray}}

\begin{document}

\title{On the structure of the vacuum state in general boundary quantum field theory}
\author{Daniele Colosi}\email{colosi@matmor.unam.mx}
\affiliation{Instituto de Matem\'aticas, Universidad Nacional Aut\'onoma
  de M\'exico, Campus Morelia,
C.P.~58190, Morelia, Michoac\'an, Mexico}
\date{\today}
\preprint{UNAM-IM-MOR-2009-1}

\begin{abstract}
We quantize a real massive Klein-Gordon field in curved spacetimes within the general boundary formulation. The vacuum wave function is given by a Gaussian in the Schr\"odinger representation and we study the general structure of the operator appearing in the Gaussian. We show that it obeys a Riccati equation and we provide the general solution.
\end{abstract}

\maketitle

\section{Introduction}
\label{sec:hyp}

This paper presents a derivation of the general structure of the vacuum state for the quantum theory of a massive Klein-Gordon field in certain curved spacetimes\footnote{In particular, the result presented here will be valid for globally hyperbolic spactimes.} within the general boundary formulation (GBF) \cite{Oe:boundary,Oe:GBQFT,Oe:KGtl}. The GBF is a new way to describe dynamical quantum fields that takes explicit account of the (properties of the) spacetime region where dynamics takes place. In particular the novelty consists in avoiding the restriction to special classes of regions and boundary hypersurfaces: in the GBF \textit{arbitrary} hypersurfaces are admissible, while in Minkowski based quantum field theory (QFT) these hypersurfaces are usually spacelike hyperplanes
. Hence within the GBF dynamics can be consistently studied even in situation that radically departs from the standard settings considered in QFT textbooks. In particular, examples of regions with connected (but non-compact) boundary as well as compact regions were investigated, \cite{Oe:KGtl,CoOe:smatrix2d}, and perturbative interacting QFT was treated for such regions in \cite{CoOe:spsmatrix,CoOe:smatrixgbf,CoOe:smatrix2d}.

In the GBF amplitudes are associated with spacetime regions and Hilbert spaces with their boundaries. The existence of one distinguished vacuum state in each of these Hilbert spaces is postulated and multiparticle states are then constructed from this vacuum state. Here we study the structure of the vacuum state for a massive scalar field in curved spacetimes and derive a general expression for it. 
The equation of motion, the Klein-Gordon equation, is a partial differential equation involving several (depending on the dimensionality of spacetime) independent variables. We will be interested only in those situations where the solutions are obtainable by the method of separation of variables. Although this technique does not have universal validity, it can be applied in most of the physically interesting cases.
We also consider a particular class of boundaries: Those that can be labeled by a constant value of one independent coordinate. This is the case, for example, of the standard situation in Minkowski spacetime where the boundaries are spacelike hypersurfaces of equal time. This requirement for the boundary is certainly a restrictive condition, however the cases studied so far in the literature belong to this category. The treatment of general boundaries will be explored elsewhere.

The paper is organized as follows. In Section \ref{sec:action} the classical theory of a massive Klein-Gordon field is considered and its action in the spacetime region of interest expressed in terms of the configurations of the field on the boundary. The quantization is performed in Section \ref{sec:pathint}, where the path integral prescription is formally implemented, and the functional Schr\"odinger representation for states is used. Section \ref{sec:vacuumstate} contains the main result of this work, namely the general structure of the vacuum wave function. The result is re-derived from a canonical treatment in Section \ref{sec:canonical}. Examples are presented in Section \ref{sec:examples}. Finally, we conclude with a recapitulation in Section \ref{sec:conclusion}.

\section{Classical theory}
\label{sec:action}
Consider the free theory of a Klein-Gordon field $\phi$ with mass $m$ propagating on a four-dimensional spacetime with line element given by
\be
\xd s^2 = g_{\mu \nu} \xd x^{\mu} \xd x^{\nu}, \qquad \mu, \nu = 0,1,2,3.
\ee 
The action in a spacetime region $M$ is
\be
S_M(\phi) = \frac{1}{2} \int_M \xd^4 x \sqrt{|g|}\left( g^{\mu \nu} \partial_{\mu} \phi \, \partial_{\nu} \phi - m^2 \phi^2 \right),
\label{eq:action}
\ee
where the integration is extended over the region $M$ and we use the notation $\partial_{\mu}= \partial / \partial x^{\mu}$, and $g \equiv \det g_{\mu \nu}$. The equation of motion is obtained by varying the action with respect to the field and setting the variation equal to zero, yielding the Klein-Gordon equation
\be
\left(\frac{1}{\sqrt{|g|}} \partial_{\mu} \left(\sqrt{|g|} g^{\mu \nu} \partial_{\nu} \right) + m^2 \right) \phi(x)=0.
\label{eq:KG}
\ee
The action for a solution of the Klein-Gordon equation can be computed performing an integration by parts in (\ref{eq:action}) and using equation (\ref{eq:KG}), yielding
\be
S_M(\phi) = \frac{1}{2} \int_{\partial M} \xd \Sigma_{\mu} \phi \left(  g^{\mu \nu}  \partial_{\nu} \phi \right),
\ee
where $\xd \Sigma_{\mu} = \xd^3 s \, \sqrt{|g^{(3)}|} \, n_{\mu}$, the coordinates on the boundary $\partial M$ are $s=(s^i), i=1,2,3$, $g^{(3)}$ is the determinant of the induced metric on the boundary and $n_{\mu}$ is the outward normal to $\partial M$. 

Consider a coordinate system $(t,\underline{x})$\footnote{We use the collective notation $\underline{x}$ to indicate the coordinates on the hypersurfaces of constant $t$.} in which the Klein-Gordon equation (\ref{eq:KG}) may be solved by the method of separation of variables. Then the solution can be written in the form
\be
\phi(t,\underline{x}) = X_{1,k}(t)Y_1(\underline{x}) + X_{2,k}(t)Y_2(\underline{x}) ,
\ee
where $X_{i,k}(t)$ is to be understood as an operator defined through its eigenvalues on a mode decomposition of $Y_i(\underline{x})$ on the hypersurface of constant $t$. We consider a spacetime region bounded by the disjoint union of two hypersurfaces, namely $\partial M = \Sigma \cup \hat{\Sigma}$ where the two hypersurfaces are defined as $\Sigma : \{ t = \xi \}$ and $\hat{\Sigma} : \{ t = \hat{\xi} \}$. The boundary configurations of the field are
\be
\phi(t,\underline{x}) \big|_{\Sigma} = \varphi(\underline{x}), \qquad \phi(t,\underline{x}) \big|_{\hat{\Sigma}} = \hat{\varphi}(\underline{x}).
\ee
We can now express the classical solution in terms of the boundary configurations $\varphi$ and $\hat{\varphi}$,
\be
\phi(t,\underline{x}) 
= \frac{\delta_k(t, \hat{\xi})}{\delta_k(\xi, \hat{\xi})} \varphi(\underline{x}) +\frac{\delta_k(\xi, t)}{\delta_k(\xi, \hat{\xi})} \hat{\varphi}(\underline{x}),
\ee
where $\delta_k(\xi, \hat{\xi}) := X_{1,k}(\xi) X_{2,k}(\hat{\xi}) - X_{1,k}(\hat{\xi}) X_{2,k}(\xi)$ is to be understood as an operator defined through its eigenvalues on a mode decomposition of the boundary configurations $\varphi(\underline{x})$ on $\Sigma$ and $\hat{\varphi}(\underline{x})$ on $\hat{\Sigma}$.
Then, in terms of these boundary configurations the action for a classical solution of the equation of motion takes the form
\be
S_M(\phi)
=  \frac{1}{2} \int_{\partial M} \xd^3 \underline{x} \, 
\begin{pmatrix}\varphi & \hat{\varphi} \end{pmatrix} W_{M}  
\begin{pmatrix} \varphi \\ \hat{\varphi} \end{pmatrix},
\ee
where the $W_M$ is a 2x2 matrix with elements $W_M^{(i,j)}, (i,j=1,2),$ given by
\bea
W_M^{(1,1)} 
&=& - \sqrt{|g^{(3)}|} \, \frac{1}{\delta_k(\xi, \hat{\xi})}\, \delta_k'(t, \hat{\xi}) \bigg|_{t \in \Sigma}, \qquad
W_M^{(1,2)} 
= - \sqrt{|g^{(3)}|} \, \frac{1}{\delta_k(\xi, \hat{\xi})} \, \delta_k'(\xi, t) \bigg|_{t \in \Sigma}, \label{eq:W1} \\
W_M^{(2,1)} 
&=& \sqrt{|\hat{g}^{(3)}|} \,  \frac{1}{\delta_k(\xi, \hat{\xi})} \, \delta_k'(t, \hat{\xi}) \bigg|_{t \in \hat{\Sigma}}, \qquad
W_M^{(2,2)} 
= \sqrt{|\hat{g}^{(3)}|}\,  \frac{1}{\delta_k(\xi, \hat{\xi})} \, \delta_k'(\xi, t) \bigg|_{t \in \hat{\Sigma}},
\label{eq:W2}
\eea
where $g^{(3)}$ and $\hat{g}^{(3)}$ are the determinants of the induced metrics on the hypersurfaces $\Sigma$ and $\hat{\Sigma}$ respectively, and a prime indicates the normal derivatives to these hypersurfaces.

\section{Path integral quantization}
\label{sec:pathint}

The passage to the quantum theory is implemented by the Feynman path integral prescription, which is the quantization procedure most suited for the GBF. Moreover, the quantum dynamics of the field is described in the Schr\"odinger representation, where the quantum states are wave functionals on the space of field configurations. 
Thus, with a given spacetime hypersurface $\Sigma$ we associate the space of state $\cH_\Sigma$ of wave functions of field configurations on $\Sigma$. This state space carries the following inner product,
\be
\langle \psi_{\Sigma}, \psi_{\Sigma}'\rangle \defeq \int \xD \varphi \, \overline{\psi_{\Sigma}(\varphi)} \, \psi_{\Sigma}'(\varphi),
\label{eq:inner-prod}
\ee
where the integral is over all field configurations $\varphi$ on the $\Sigma$.
Amplitudes $\rho_{M}:\cH_{\partial M}\to\C$ are associated to certain spacetime regions $M$. State spaces and amplitudes satisfy a number of consistency conditions, see \cite{Oe:GBQFT} or \cite{Oe:KGtl}.

The field propagator associated with the spacetime region $M$ with boundary $\partial M = \Sigma \cup \hat{\Sigma}$, is formally defined as
\be
Z_M(\varphi, \hat{\varphi}) = \int_{\phi|_{\Sigma}=\varphi, \, \phi|_{\hat{\Sigma}}=\hat{\varphi} } \xD\phi\, e^{\im S_{M}(\phi)},
\label{eq:proppint0}
\ee
where $S_M(\phi)$ is the action of the field in the region $M$ and the integration is extended over all field configurations $\phi$ that reduce to the boundary configurations $\varphi$ and $\hat{\varphi}$ on the boundary hypersurfaces $\Sigma$ and $\hat{\Sigma}$ respectively. All the information on the dynamical evolution of the field between boundary configurations $\varphi$ and $\hat{\varphi}$ is encoded in the propagator (\ref{eq:proppint0}). In the case of the free theory determined by the free action (\ref{eq:action}) we can evaluate the associated propagator by shifting the integration variable by a classical solution, $\phi_\text{cl}$, matching the boundary configurations in $\partial M = \Sigma \cup \hat{\Sigma}$, i.e. $\phi_\text{cl}|_{\Sigma} = \varphi$ and $\phi_\text{cl}|_{\hat{\Sigma}} = \hat{\varphi}$. Explicitly,
\be
Z_{M}(\varphi, \hat{\varphi}) = \int_{\phi|_{\Sigma}=\varphi, \, \phi|_{\hat{\Sigma}}=\hat{\varphi} } \xD\phi\, e^{\im S_{M}(\phi)}
=  \int_{\phi|_{\partial M}=0} \xD\phi\, e^{\im  S_{M}(\phi_{cl}+\phi)} = N_{M} \,  e^{\im  S_{M}(\phi_\text{cl})} ,
\label{eq:propfree}
\ee
where the normalization factor is formally given by
\be
N_{M}=\int_{\phi|_{\partial M}=0} \xD\phi\, e^{\im S_{M}(\phi)}.
\label{eq:n00}
\ee

\section{Vacuum state}
\label{sec:vacuumstate}

In the state space ${\cal H}_{\Sigma}$ associate to the a hypersurface $\Sigma$ we postulate the existence of a distinguished vacuum wave function. As in \cite{Oe:timelike,Oe:KGtl} we make for these vacuum wave function the Gaussian ansatz
\be
 \psi_{\Sigma,0}(\varphi) = C  \exp\left(-\frac{1}{2}\int_\Sigma \xd^3 \underline{x} \, \sqrt{|g^{(3)}|} \,\varphi(\underline{x})(A \varphi)(\underline{x})\right), 
\label{eq:vacuum}
\ee
where $C$ is a normalization factor (it can be calculated with the inner product (\ref{eq:inner-prod}) requiring the vacuum state to normalized) and $A$ is an unknown operator called the vacuum operator. 
The vacuum state is a fundamental ingredient in the construction of the quantum theory. It provides the basis for generating multi-particle states: These are given by a product of polynomials of the field configurations with the vacuum wave function. Moreover the product of the Gaussian wave function (\ref{eq:vacuum}) with its complex conjugate provides the measure on the space of field configuration\footnote{In \cite{CCQ:schroecurv,CCQ:schroefock}, the analysis of the Schr\"odinger representation for the quantum theory of a scalar field in curved spacetimes showed that this measure is indeed a Gaussian.} making the formal expression (\ref{eq:inner-prod}) well defined. 

The vacuum state is required to satisfy the vacuum axioms of \cite{Oe:GBQFT}, see also \cite{Oe:KGtl}. This implies in particular that it is invariant under free evolution, from one hypersurface to another, say from $\Sigma$ to $\hat{\Sigma}$, implemented by the action of the (free) field propagator (\ref{eq:proppint0}).
The invariance of the vacuum state reads
\be
\psi_{\Sigma,0}(\varphi) = \int \xD \hat{\varphi} \, \psi_{\hat{\Sigma},0}(\hat{\varphi})  \,Z_{M}(\varphi, \hat{\varphi}),
\ee
where the integration is over all field configurations $\hat{\varphi}$ on $\hat{\Sigma}$.
This equation implies for the vacuum operators $A$ on $\Sigma$ and $\hat{A}$ on $\hat{\Sigma}$ the relation
\be
\left(\sqrt{|g^{(3)}|}A-i W_M^{(1,1)} \right) \left(\sqrt{|\hat{g}^{(3)}|}\hat{A}+i W_M^{(2,2)} \right) = \frac{\left( W_M^{(1,2)} + W_M^{(2,1)}\right)^2}{4}
\label{eq:vo}
\ee
We now consider this equation in the case where the two surface $\Sigma$ and $\hat{\Sigma}$ are infinitesimally close to each other, so that (using the notation of Section \ref{sec:action}) we can write $\hat{\xi} \approx \xi + \xd \xi$, with $\xd \xi \ll 1$. The next step is to consider the expansion of the various terms depending on $\hat{\Sigma}$ in terms of quantities depending only on $\Sigma$ and $\xd \xi$:
\bea
\sqrt{|\hat{g}^{(3)}|} &\approx& \sqrt{|g|^{(3)}} + \left( \sqrt{|g|^{(3)}} \right)' \xd \xi + o(\xd \xi), \\
\hat{A} &\approx& A + A' \xd \xi + o(\xd \xi), \\
W_M^{(1,1)} &\approx& \frac{\sqrt{|g^{(3)}|}}{ \xd \xi} - \sqrt{|g^{(3)}|} (\ln {\cal W})' - \sqrt{|g^{(3)}|}\frac{{\cal W}_{1,2}}{{\cal W}} \xd \xi+ o(\xd \xi),\\
W_M^{(1,2)} &\approx& - \frac{\sqrt{|g^{(3)}|}}{ \xd \xi} + \sqrt{|g^{(3)}|} (\ln {\cal W})' + o(\xd \xi),\\
W_M^{(2,1)} &\approx& - \frac{\sqrt{|g^{(3)}|}}{ \xd \xi} - \left( \sqrt{|g^{(3)}|} \right)'+ \sqrt{|g^{(3)}|} \left[(\ln {\cal W})'\right]^2 \xd \xi - \sqrt{|g^{(3)}|} \frac{{\cal W}_{0,3}}{{\cal W}} \xd \xi + o(\xd \xi),\\
W_M^{(2,2)} &\approx&  \frac{\sqrt{|g^{(3)}|}}{ \xd \xi} + \left( \sqrt{|g^{(3)}|} \right)' - \sqrt{|g^{(3)}|} \left[(\ln {\cal W})'\right]^2 \xd \xi + \sqrt{|g^{(3)}|} \frac{{\cal W}_{0,3}}{{\cal W}} \xd \xi + o(\xd \xi),
\eea
where ${\cal W} = X_{1,k}(\xi) X_{2,k}'(\xi) - X_{2,k}(\xi) X_{1,k}'(\xi)$, ${\cal W}_{0,3} = X_{1,k}(\xi) X_{2,k}'''(\xi) - X_{2,k}(\xi) X_{1,k}'''(\xi)$ and ${\cal W}_{1,2} = X_{1,k}'(\xi) X_{2,k}''(\xi) - X_{2,k}'(\xi) X_{1,k}''(\xi)$. The substitution of the above expressions in equation (\ref{eq:vo}) leads (up to order 0 in $(\xd \xi)$) to the following Riccati equation
\be
A'+ a_1 A^2 + a_2 A + a_3 =0,
\label{eq:Riccati}
\ee
where the quantities $a_i$ are given by
\be
a_1 = \im, \qquad a_2= - (\ln {\cal W})', \qquad a_3 = - \im \frac{{\cal W}_{1,2}}{{\cal W}} - \frac{\im }{4 g^{(3)} {\cal W}^2} \left( \left[\sqrt{|g^{(3)}|}{\cal W}\right]' \right)^2.
\ee

Multiplying the term $\sqrt{|g^{(3)}|}{\cal W}$ by $Y_1(\underline{x})Y_2(\underline{x})$ and integrating over the surface $\Sigma$, leads to the inner product on the space of classical solution of the equation of motion,
\be
\int_{\Sigma} \xd^3 \underline{x} \sqrt{|g^{(3)}|} \left(\phi_1 \phi_2'- \phi_1' \phi_2 \right),
\label{eq:innerprod}
\ee
where $\phi_1$ and $\phi_2$ are two solutions of the equation of motion. This inner product is independent of the surface $\Sigma$. We note that all the dependence on the surface $\Sigma$ is contained on the term $\sqrt{|g^{(3)}|}{\cal W}$, and therefore we can conclude that this term is a constant and its derivative vanishes. Hence the Riccati equation (\ref{eq:Riccati}) results to be
\be
A' + \im A^2 - (\ln {\cal W})'A - \im \frac{{\cal W}_{1,2}}{{\cal W}} =0.
\ee
Writing the operator $A= - \im u'/u$, the quantity $u$ satisfies the equation
\be
u'' - (\ln {\cal W})' u' + \frac{{\cal W}_{1,2}}{{\cal W}} u=0.
\ee
It is straightforward to verify that any linear combination of $X_{1,k}$ and $X_{2,k}$ solves the above equation. Hence $u= c_1 X_{1,k} + c_2 X_{2,k}$, where $c_i$ are constants. Therefore the general form of the vacuum operator is
\be
A= - \im \left[ \ln (c_1 X_{1,k} + c_2 X_{2,k}) \right]'.
\label{eq:Asol}
\ee
It can be verified, with simple algebra, that the above expression for the operator $A$ satisfies equation (\ref{eq:vo}).

\section{Canonical treatment}
\label{sec:canonical}

In this section we will recover the result (\ref{eq:Asol}) via a canonical treatment for the scalar field considered in the Schr\"odinger representation.

The usual way to canonically quantize a field is to impose canonical commutation relations to the field $\phi$ and its conjugate momentum $\pi$. Hence we define the conjugate momentum to $\phi$,
\be
\pi= \frac{\partial {\cal L}}{\partial (n^{\mu} \partial_{\mu} \phi)},
\ee
where $\cal L$ is the Lagrangian ,its expression can be read from the action (\ref{eq:action}). Then we regard $\phi$ and $\pi$ as operators (operator-valued distributions) and postulate canonical commutation relations on the surface $\Sigma: \{t= const \}$,
\be
[\phi(x), \pi(x')]\big|_{\Sigma} =  \im \delta(\underline{x}-\underline{x}'),
\label{eq:CCR}
\ee
and all the other commutators vanish. These commutation relations can be realized by representing the operators $\phi$ and $\pi$, when acting on functionals $\Psi[\varphi]$, as
\be
\phi \Psi[\varphi] = \varphi \, \Psi[\varphi], \qquad
\pi \Psi[\varphi] = - \im \frac{\delta}{\delta \varphi} \Psi[\varphi].
\ee

Now we want to expand the field in terms of creation and annihilation operators. To achieve this, we choose a basis for which the inner product defined in (\ref{eq:innerprod}) takes a simple form
\be
(P_i,Q_j)= \delta_{ij}, \qquad (P_i,P_j) = (Q_i,Q_j)=0,
\label{eq:ip2}
\ee
where $P_i$ and $Q_i$ are solutions of the equation of motion. In particular they can be written as linear combinations of $\phi_1$ and $\phi_2$. The scalar field can then be expanded in this basis,
\be
\phi = \int \xd k \, \left( a_k P_k + a_k^{\dagger} Q_k \right).
\ee
If the parameter $k$ turns out to be discrete, the integral must be replaced by a sum. The coefficients $a_k$ and $a_k^{\dagger}$ are now promoted to annihilation and creation operators respectively, and we postulate the following commutation relations
\be
[a_k,a_j^{\dagger}] = \delta_{kj},
\ee
in order to reproduce the relations (\ref{eq:CCR}). We associate a Hilbert space $\cal H$ to the quantization surface $\Sigma$. The vacuum state $\Psi_0 \in \cal H$ is defined by the equation
\be
a_k \Psi_0 =0.
\label{eq:vac}
\ee
The annihilation operator can be expressed in terms of the field and the conjugate momentum with the inner product (\ref{eq:ip2}),
\be
a_n = - (Q_k, \phi) = \int_{\Sigma} \xd^3 \underline{x} \, \left(\phi \sqrt{|g^{(3)}|} Q_k' - Q_k \pi  \right).
\label{eq:annihil}
\ee
The substitution of expressions (\ref{eq:annihil}) and (\ref{eq:vacuum}) in equation (\ref{eq:vac}) leads to the equation
\be
\int_{\Sigma} \xd^3 \underline{x} \, \sqrt{|g^{(3)}|} \, \left[ Q_k' - \im Q_k A \right]\varphi(y) = 0,
\ee
which fixes the expression of $A$,
\be
A= - \im \left[ \ln Q_k \right]'.
\ee
This coincides with what has been found in the previous section, namely equation (\ref{eq:Asol}).

\section{Examples}
\label{sec:examples}

As explicit examples of the result presented in the preceding sections, we will provide the expression of the vacuum wave function for the Klein-Gordon field in Rindler and de Sitter spacetimes. Before dealing with the mentioned spacetimes we notice that the scalar field in Minkowski space has already been studied in literature, where different spacetime regions with different boundary hypersurfaces have been considered. In particular in \cite{Oe:KGtl} state spaces were associated with spacelike hyperplanes, timelike hyperplanes and a timelike hypercylinder, namely a sphere in space extended over all of time, i.e. $ \mathbb{R} \times S^2$ (see in particular formula (80) of \cite{CoOe:smatrixgbf}), and the vacuum state there defined can be expressed according to formula (\ref{eq:Asol}). Finally, for compact circular region in two dimensional Euclidean spacetime formula (\ref{eq:Asol}) provides the correct vacuum state operator \cite{CoOe:smatrix2d}. Let us now consider spacetimes different from the Minkowski one.

\subsection{Rindler spacetime}

We consider a real massive scalar field in 4 dimensional Rindler spacetime, the geometry of which is described by the metric
\be
\xd s^2 = \rho^2 \xd \eta^2 - \xd \rho^2 -\xd y^2 - \xd z^2, \qquad -\infty < \eta < + \infty, \qquad 0< \rho< \infty.
\ee
The Klein-Gordon equation (\ref{eq:KG}) in Rindler space takes the form 
\be
\left( \frac{1}{\rho^2}\partial_{\eta}^2 - \frac{1}{\rho} \partial_{\rho} \rho \partial_{\rho} - \partial_y^2-\partial_z^2 + m^2\right) \phi(\eta,\rho,y,z) =0,
\ee
and the general solution can be written as
\be
\phi(\eta,\rho,y,z) = \int_0^{\infty} \xd \mu \int_{-\infty}^{\infty} \xd q_y \int_{-\infty}^{\infty} \xd q_z \, \left( f_{\mu}(\rho) e^{\im \mu \eta} e^{\im (q_y y+ q_z z)} + \overline{f_{\mu}}(\rho) e^{-\im \mu \eta} e^{-\im (q_y y + q_z z)}\right),
\label{eq:fourier}
\ee
where $f_{\mu}(\rho)$ are given in terms of Bessel functions,
\be
f_{\mu}(\rho) = a_{k} I_{\im \mu}(k \rho) + b_{k} K_{\im \mu}(k \rho),
\ee
with $k = \sqrt{m^2+q_y^2+q_z^2}$. $I_{\im \mu}$ and $K_{\im \mu}$ are the modified Bessel functions of the first and second kind respectively, of imaginary order $\im \mu$. They represent two linearly independent solutions of the modified Bessel equation; their Wronskian results to be \cite{Wat:Bessel},
\be
{\cal W}\{K_{\nu}(z), I_{\nu}(z) \}= \frac{1}{z}.
\label{eq:wronIK}
\ee
Notice that both these Bessel functions have an oscillatory behavior in a neighborhood of the origin $(\rho=0)$, \cite{GR:tables},
\be
I_{\im \mu}(z) \approx \left( \frac{z}{2} \right)^{\im \mu}/ \Gamma(\im \mu +1), \qquad K_{\im \mu}(z) \approx \frac{1}{2}\left( \frac{z}{2} \right)^{-\im \mu} \Gamma(\im \mu).
\ee
On the other hand they behave very differently for large values of their argument \cite{GR:tables},
\be
I_{\im \mu}(z) \approx \frac{e^z}{\sqrt{2 \pi z}}, \qquad K_{\im \mu}(z) \approx \sqrt{\frac{\pi}{2 z}} \, e^{-z}.
\ee
These asymptotic behaviors allow to select the appropriate solution depending on the spacetime region of interest. 	
%
In particular, we will define the vacuum state for two different types of hypersurfaces, the hypersurfaces defined by a constant value of the coordinates $\rho$ and $\eta$ respectively.

\subsubsection{Hypersurface $\rho = const$} 

We will be interested in studying the field in an infinite region bounded by two hypersurfaces of constant $\rho$. The determinant of the 3-metric induced on these hypersurfaces turns out to be $g^{(3)} = \rho^2$, and the normal derivative is $\partial_{\rho}$.
In the specified region the classical solution may contain both kinds of Bessel functions because both $I_{\im \mu}$ and $K_{\im \mu}$ are regular there. Hence the classical solution can be decomposed formally in the form
\be
\phi(\eta, \rho, y,z) = I_{\im \mu}(k \rho) \, \varphi_{I}(\eta,y,z) + K_{\im \mu}(k \rho) \, \varphi_{K}(\eta,y,z),
\label{eq:div+conv}
\ee
where $\varphi_I$ and $\varphi_K$ are real functions on the hypersurface $\rho=const$. $I_{\im \mu}(k \rho)$ and $K_{\im \mu}(k \rho)$ are understood as operators, defined through their eigenvalues on a Fourier decomposition of $\varphi_{I}$ and $\varphi_{K}$ on the hypersurface $\rho=const$. Denoting with $\varphi$ and $\hat{\varphi}$ the configurations on the boundaries of the region specified by the hypersurfaces labeled by coordinates $\rho=\xi$ and $\rho= \hat{\xi}$, with $\hat{\xi} > \xi$, respectively, we express the classical solution (\ref{eq:div+conv}) in terms of the boundary configurations,
\be
\phi(\eta, \rho, y,z) = \frac{\delta_{\mu} (k \rho,k \hat{\xi})}{\delta_{\mu} (k \xi,k \hat{\xi})} \, \varphi(\eta,y,z) + \frac{\delta_{\mu} (k\xi,k \rho)}{\delta_{\mu} (k \xi,k \hat{\xi})} \, \hat{\varphi}(\eta,y,z),
\ee
where 
\be
\delta_{\mu}(z,\hat{z}) \defeq I_{\im \mu}(z) \, K_{\im \mu}(\hat{z}) - K_{\im \mu}(z) \, I_{\im \mu}(\hat{z}),
\label{eq:delta}
\ee
again $\delta_{\mu}$ is to be understood as an operator acting on ${\varphi}$ and $\hat{\varphi}$.
We can now express the quantities in (\ref{eq:W1}-\ref{eq:W2}) as
\be
W_{M}^{(1,1)} = \xi  k \frac{ \sigma_{\mu}(k \hat{\xi}, k \xi)}{\delta_{\mu}(k \xi, k \hat{\xi})}, \qquad W_{M}^{(1,2)} =W_{M}^{(2,1)} = \frac{1}{\delta_{\mu}(k \xi, k \hat{\xi})}, \qquad W_{M}^{(2,2)} = \hat{\xi} k  \frac{\sigma_{\mu}(k \xi, k \hat{\xi})}{\delta_{\mu}(k \xi, k \hat{\xi})},
\ee
where the Wronskian relation (\ref{eq:wronIK}) has been used.
The function $\sigma_{\mu}$ is to be understood as the operator defined as
\begin{gather*}
\sigma_{\mu}(\hat{z},z)  = I_{\im \mu}(\hat{z}) \, K_{\im \mu}'(m z) - I_{\im \mu}'(z) \, K_{\im \mu}(\hat{z}),
\end{gather*}
where the prime here indicates a derivative with respect to $z$.
The equation (\ref{eq:vo}) that defines the vacuum operator takes the form
\be
\xi \, \hat{\xi} \left( A - \im k \frac{ \sigma_{\mu}(k \hat{\xi}, k \xi)}{\delta_{\mu}(k \xi, k \hat{\xi})} \right)\left( \hat{A} + \im k \frac{ \sigma_{\mu}(k \xi, k \hat{\xi})}{\delta_{\mu}(k \xi, k \hat{\xi})} \right) = \frac{1}{\delta_{\mu}(k \xi, k \hat{\xi})^2}.
\ee
It can be easily verified using the Wronskian relation (\ref{eq:wronIK}) that expression (\ref{eq:Asol}), written in the form $A = - \im \partial_n \left[ \ln (c_1 I_{\im \mu}(k z) + c_2 K_{\im \mu}(k z))\right]$, where $\partial_n$ denotes the normal derivative to the hypersurface $\Sigma$ of constant $\rho$, solves this equation. Finally, in order to have a well defined vacuum state, we require that the argument of the exponential (\ref{eq:vacuum}) be bounded from below. We can select for the constants $c_1$ and $c_2$ the values -1 and 0 respectively. Hence, the vacuum state defined on $\Sigma$ reads
\be
 \psi_{\Sigma,0}(\varphi) = C  \exp\left(\frac{\im}{2}\int_\Sigma \xd \eta \, \xd y \, \xd z \, \rho \, k\, \varphi(\eta,y,z)\frac{I_{\im \mu}'(k \rho)}{I_{\im \mu}(k \rho)} \varphi(\eta,y,z)\right).
\ee

\subsubsection{Hypersurface $\eta = const$} 

We define in this section the vacuum state on hypersurfaces of constant $\eta$. Then we consider the spacetime region $M$ bounded by two hypersurfaces $\Sigma:=\{ \eta= \xi\}$ and $\hat{\Sigma}:=\{\eta =\hat{\xi} \}$. The determinant of the 3-metric induced on these hypersurfaces turns out to be 1, and the normal derivative is $(1/\rho) \partial_{\eta}$. In the specified region the classical solution of the Klein-Gordon equation can be formally written as 
\be
\phi(\eta, \rho, y,z) = e^{-\im \mu \eta} \, \varphi^+(\rho,y,z) + e^{\im \mu \eta}  \, \varphi^-(\rho,y,z),
\ee
where $\mu$ is understood as the operator of the form $\mu = \sqrt{- \rho \partial_{\rho} \rho \partial_{\rho} - \rho^2 ( \partial_y^2+\partial_z^2 - m^2)}$. Denoting the boundary configurations by $\varphi$ on $\Sigma$ and by $\hat{\varphi}$ on $\hat{\Sigma}$, we can expand the classical solution in terms of these configurations,
\be
\phi(\eta, \rho, y,z) = \frac{\sin (\mu( \hat{\xi} - \eta))}{\sin (\mu( \hat{\xi} - \xi))} \, \varphi(\rho,y,z) + \frac{\sin (\mu( \eta- \xi))}{\sin (\mu( \hat{\xi} - \xi))} \, \hat{\varphi}(\rho,y,z),
\ee
where the fractions are understood as operators. We obtain for the quantities (\ref{eq:W1}-\ref{eq:W2}) the expressions,
\be
W_{M}^{(1,1)} = W_{M}^{(2,2)} = \frac{\mu}{\rho}\, \frac{\cos (\mu( \hat{\xi} - \xi))}{\sin (\mu( \hat{\xi} - \xi))}, \qquad W_{M}^{(1,2)} =W_{M}^{(2,1)} = - \frac{\mu}{\rho} \frac{1}{\sin (\mu( \hat{\xi} - \xi))},
\ee
and equation (\ref{eq:vo}) reads
\be
\left( A - \im \frac{\mu}{\rho}\, \frac{\cos (\mu( \hat{\xi} - \xi))}{\sin (\mu( \hat{\xi} - \xi))} \right)\left( \hat{A} + \im \frac{\mu}{\rho}\, \frac{\cos (\mu( \hat{\xi} - \xi))}{\sin (\mu( \hat{\xi} - \xi))} \right) = \frac{\mu^2}{\rho^2} \frac{1}{\sin^2 (\mu( \hat{\xi} - \xi))}.
\ee
The solution coincides with expression (\ref{eq:Asol}), in the form $A= - \im \partial_n \left[\ln \left(c_1 e^{\im \mu z} + c_2 e^{-\im \mu z} \right) \right]$, with $\partial_n$ indicating the normal derivative to the hypersurface of constant $\eta$. An appropriate choice for the coefficients $c_1$ and $c_2$ in order to define the vacuum state is $c_1=1$ and $c_2=0$. Hence we arrive at the expression of the vacuum state of an hypersurface of constant $\eta$,
\be
 \psi_{\Sigma,0}(\varphi) = C  \exp\left(-\frac{1}{2}\int_\Sigma \frac{\xd \rho \, \xd y \, \xd z}{\rho} \,  \varphi(\rho,y,z) \, \mu \, \varphi(\rho,y,z)\right).
\ee

\subsection{de Sitter spacetime}

We use the coordinate system in which the de Sitter metric has the form (t>0),
\bea
ds^2 &=& \frac{R^2}{t^2} \left( \xd t^2 - (\xd x)^2 - (\xd y)^2- (\xd z)^2 \right), \label{eq:dSmetric} \\
&=& \frac{R^2}{t^2} \left( \xd t^2 - (\xd \underline{x})^2  \right).
\eea
The Klein-Gordon equation (\ref{eq:KG}) in de Sitter space,
\be
\left[ \frac{t^2}{R^2} \left(\partial_t^2 - \partial_x^2 - \partial_y^2 - \partial_z^2 \right) -\frac{2t}{R^2} \partial_t +m^2 \right]\phi(t, \underline{x}) = 0,
\ee
has the following general solution,
\be
\phi(t, \underline{x}) =\int \frac{\xd ^3 \underline{k}}{(2 \pi)^{3/2}} \left( v_k(t) \, e^{\im \underline{k} \cdot \underline{x}} +  \overline{v_k}(t) \, e^{-\im \underline{k} \cdot \underline{x}} \right),
\ee
where
\be
v_k(t)= t^{3/2} \left(c_{1, k} J_{\nu}(kt) + c_{2,k} Y_{\nu}(kt) \right),
\ee
where $k=|\underline{k}|$, and $J_{\nu}(z)$ and $Y_{\nu}(z)$  are the Bessel functions of the first and second kind respectively, with index $\nu = \sqrt{\frac{9}{4} - (m R)^2}$. 
We consider a spacetime region bounded by two hypersurfaces of constant $t$, namely $\Sigma := \{t= \xi \}$ and $\hat{\Sigma} := \{t= \hat{\xi} \}$, and assume $\hat{\xi}> \xi$.  The 3-metric induced on the hypersurface of constant $t$ corresponds to the spatial part of the metric (\ref{eq:dSmetric}), and its determinant is $g^{(3)}= (R/t)^6 $. The normal derivative to this hypersurface results to be $(t/R)\partial_t$. 

As in the preceeding sections the field configurations on the boundary are indicated by $\varphi$ and $\hat{\varphi}$ on $\Sigma$ and $\hat{\Sigma}$ respectively. Treating the Bessel functions as operators, we can express the classical solution of the Klein-Gordon equation in terms of the boundary configurations as 
\be
\phi(t, \underline{x}) = \frac{\delta (k t,k \hat{\xi})}{\delta (k \xi,k \hat{\xi})} \, \varphi(\underline{x}) + \frac{\delta (k\xi,k t)}{\delta (k \xi,k \hat{\xi})} \, \hat{\varphi}(\underline{x}),
\ee
where 
\be
\delta(k z,k \hat{z}) \defeq z^{3/2} \, \hat{z}^{3/2} \left[ J_{\nu}(k z) \, Y_{\nu}(k \hat{z}) - Y_{\nu}(k z) \, J_{\nu}(k \hat{z})\right].
\label{eq:delta2}
\ee
Then the quantities in (\ref{eq:W1}-\ref{eq:W2}) read
\bea
W_M^{(1,1)} &=& - \frac{R^2}{\xi^2} \left( \frac{3}{2}\frac{1}{\xi} + k \frac{J_{\nu}'(k \xi) \, Y_{\nu}(k \hat{\xi}) - Y_{\nu}'(k \xi) \, J_{\nu}(k \hat{\xi})}{J_{\nu}(k \xi) \, Y_{\nu}(k \hat{\xi}) - Y_{\nu}(k \xi) \, J_{\nu}(k \hat{\xi})}\right) ,\nonumber\\
W_M^{(1,2)} &=& W_M^{(2,1)} = - \frac{2 R^2}{\pi \delta (k \xi,k \hat{\xi})},\nonumber\\
W_M^{(2,2)} &=& \frac{R^2}{\hat{\xi}^2} \left( \frac{3}{2}\frac{1}{\hat{\xi}} + k \frac{J_{\nu}(k \xi) \, Y_{\nu}'(k \hat{\xi}) - Y_{\nu}(k \xi) \, J_{\nu}'(k \hat{\xi})}{J_{\nu}(k \xi) \, Y_{\nu}(k \hat{\xi}) - Y_{\nu}(k \xi) \, J_{\nu}(k \hat{\xi})}\right),
\eea
where the prime denotes the derivative with respect to the argument.
Substituting these quantities in expression (\ref{eq:vo}) we obtain
\bea
&\left[\frac{R^3}{\xi^3} A + \im \frac{R^2}{\xi^2} \left( \frac{3}{2}\frac{1}{\xi} + k \frac{J_{\nu}'(k \xi) \, Y_{\nu}(k \hat{\xi}) - Y_{\nu}'(k \xi) \, J_{\nu}(k \hat{\xi})}{J_{\nu}(k \xi) \, Y_{\nu}(k \hat{\xi}) - Y_{\nu}(k \xi) \, J_{\nu}(k \hat{\xi})}\right) \right] 
\left[\frac{R^3}{\hat{\xi}^3} \hat{A} + \im \frac{R^2}{\hat{\xi}^2} \left( \frac{3}{2}\frac{1}{\hat{\xi}} + k \frac{J_{\nu}(k \xi) \, Y_{\nu}'(k \hat{\xi}) - Y_{\nu}(k \xi) \, J_{\nu}'(k \hat{\xi})}{J_{\nu}(k \xi) \, Y_{\nu}(k \hat{\xi}) - Y_{\nu}(k \xi) \, J_{\nu}(k \hat{\xi})}\right) \right] \nonumber\\
&= \frac{4 R^4}{\pi^2 \delta^2 (k \xi,k \hat{\xi})}.
\eea
With expression (\ref{eq:Asol}) written as $A= - \im \partial_n \left[ \ln (z^{3/2}(c_1 J_{\nu}(k z) + c_2 Y_{\nu}(k z)))\right]$ and using the Wronskian between the Bessel functions, $W(J_{\nu}(z),Y_{\nu}(z)) = 2/(\pi z)$, we can reduce the above equation to an identity. Hence the vacuum operator solution to (\ref{eq:vo}) is (\ref{eq:Asol}). Again, we select a specific linear combination of Bessel functions in order to have a well defined vacuum state, in particular we choose $c_1=1$ and $c_2= \im$. Then, the vacuum state defined on an hypersurfaces $\Sigma$ of constant $t$ takes the form
\be
 \psi_{\Sigma,0}(\varphi) = C  \exp\left(\frac{\im}{2}\int_\Sigma \xd^3 \underline{x} \, \frac{R^2}{t^2} \,  \varphi(\underline{x})\left[ k \frac{H_{ \nu}'(k t)}{H_{\nu}(k t)} + \frac{3}{2 t}\right] \varphi(\underline{x})\right).
\ee 

\section{Summary}
\label{sec:conclusion}

The aim of this paper was to present an analysis of the general structure of the vacuum wave function for a quantum Klein-Gordon field in curved spacetime within the GBF. We have implemented a path integral quantization of the field and studied the free evolution of the vacuum state between infinitesimally close hypersurfaces. This allows us to show that the vacuum operator, i.e. the operator appearing in the Gaussian of the vacuum wave function, obeys a Riccati equation, and we provided the general solution. This result has been subsequently recovered from a canonical treatment of the quantum field. 

The examples presented in the last section not only provide a confirmation of the result obtained but also represent the first application of the general boundary formulation for quantum field theory in spacetimes different from Minkowski or Euclidean spacetimes. Apart from the relevance for the GBF program in general, the present result contributes also to the construction of the Schr\"odinger representation for quantum fields on curved spacetimes. Such representation has received attention (see in particular \cite{CCQ:schroecurv,CCQ:schroefock} and references therein) motivated by its application to canonical quantum gravity, as well as to some symmetry reduced gravitational systems, such as the Gowdy $T^3$ model  \cite{Corichi:2007ht}. 

Finally, we notice that our treatment is based on a specific assumption for the hypersurfaces on which the vacuum state has been constructed: We limit our analysis to constant coordinate hypersurfaces and evolution from one such hypersurface to another. The next step will be to consider hypersurfaces of arbitrary shape as well as evolution implemented by arbitrary local deformations of the hypersurfaces.










\begin{acknowledgments}

I am grateful to Robert Oeckl for many helpful discussions and comments on an early draft of this paper. This work was supported in part by CONACyT grants 49093.

\end{acknowledgments}

\bibliographystyle{amsordx}
\bibliography{stdrefs,refs2}

\end{document}